\documentstyle[12pt,epsf]{article}
\def \bep{\mbox{\boldmath $\epsilon$}}
\def \beq{\begin{equation}}
\def \bph{{\bf \hat p}}
\def \eeq{\end{equation}}

\begin{document}
\rightline{EFI-95-65}
\rightline{FERMILAB-PUB-95/345-T}
\rightline{ANL-HEP-PR-95-80}
\rightline{WIS-95/50-Oct.-Ph}
\rightline{hep-ph/9511363}
\rightline{November 1995}
\bigskip
\bigskip
\centerline{\bf ANGULAR DISTRIBUTIONS AND LIFETIME DIFFERENCES}
\centerline{{\bf IN $B_s \to J/\psi \phi$ DECAYS}
\footnote{To be submitted to Physics Letters B}}
\bigskip
\centerline{{\it Amol S. Dighe}
\footnote{Enrico Fermi Institute and Department of Physics,
University of Chicago, Chicago, IL 60637}}
\medskip
\centerline{{\it Isard Dunietz}
\footnote{Theoretical Physics Division, Fermi National Accelerator
Laboratory, Batavia, IL 60510}}
\medskip
\centerline{{\it Harry J. Lipkin}
\footnote{High Energy
Physics Division, Argonne National Laboratory, Argonne, IL 60439}$^,$
\footnote{Department of Particle Physics,
Weizmann Institute of Science, Rehovoth, Israel}}
\medskip
\centerline{and}
\medskip
\centerline{{\it Jonathan L. Rosner}$^{~2,~3}$}
\bigskip
\centerline{\bf ABSTRACT}
\medskip
\begin{quote}
The strange $B$ meson $B_s \equiv \bar b s$ and its charge-conjugate $\bar B_s
\equiv b \bar s$ are expected to mix with one another in such a way that the
mass eigenstates $B_s^H$ (``heavy'') and $B_s^L$ (``light'') may have a
perceptible
lifetime difference of up to 40\%, with the CP-even eigenstate being
shorter-lived.  A simple transversity analysis permits one to separate the
CP-even and CP-odd components of $B_s \to J/\psi \phi$, and thus to determine
the lifetime difference.  The utility of a similar analysis for $B^0 \to J/\psi
K^{*0}$ is noted.
\end{quote}
\bigskip

The Cabibbo-Kobayashi-Maskawa picture of weak charge-changing transitions
\cite{CKM} predicts the strange $B$ meson $B_s \equiv \bar b s$ and its
charge-conjugate $\bar B_s \equiv b \bar s$ to mix with one another with a
large amplitude. The mass eigenstates $B_s^H$ (``heavy'') and $B_s^L$
(``light'') with masses $m(B_s^H) \equiv m_H$ and $m(B_s^L) \equiv m_L$ are
expected to be split by $\Delta m \equiv m_H - m_L \approx 25 \bar \Gamma$,
give or take a factor of two \cite{JRCP}, where $\bar \Gamma \equiv (\Gamma_H +
\Gamma_L)/2 \approx \Gamma(B^0)$ $(B^0 \equiv \bar b d)$ and $\Gamma_{H,L}
\equiv \Gamma(B_s^H,B_s^L)$.  The measurement of such a large mass difference
poses an experimental challenge.

To a good approximation, CP violation can be neglected in calculating the mass
eigenstates, in which case they correspond to those $B_s^{(\pm)}$ of even and
odd CP, with $B_s^L = B_s^{(+)}$ and $B_s^H = B_s^{(-)}$ as we shall see.
The decay of a $\bar B_s$ meson via the quark subprocess $b (\bar s)
\to c \bar c s (\bar s)$ gives rise to predominantly CP-even final states
\cite{CPeven}. Thus the CP-even eigenstate should have the greater decay rate.
An explicit calculation \cite{Blifes} gives
\beq \label{eqn:widthdiff}
\frac{\Gamma(B_s^{(+)}) - \Gamma(B_s^{(-)}) }{\overline \Gamma} \simeq 0.18
\frac{f_{B_s}^2}{(200~{\rm MeV})^2}~~~,
\eeq
where $f_{B_s}$ is the $B_s$ decay constant (in a normalization in which $f_\pi
= 132$ MeV).  In one estimate \cite{JRCP}, $f_{B_s} = 225 \pm 40$ MeV, while
a compilation of lattice results \cite{Soni} obtains $f_{B_s} = 201 \pm 40$ MeV
(90\% c.l. limits).  The
upper limit of 40\% for (\ref{eqn:widthdiff}) is based on an estimate of the
maximum possible contribution from the $b (\bar s) \to c \bar c s (\bar s)$
subprocess \cite{IsiBs,BDY}.

The ratio of the mass splitting to the width difference of strange $B$'s is
predicted to be large and independent of CKM matrix elements \cite{IsiBs,BP}
(to lowest order, neglecting QCD corrections which may be appreciable):
\beq
\frac{\Delta m}{\Delta \Gamma} \simeq - \frac{2}{3 \pi} \frac{m_t^2 h(m_t^2
/M_W^2)}{m_b^2} \left( 1 - \frac{8}{3} \frac{m_c^2}{m_b^2} \right)^{-1}
\simeq - 200~~~,
\eeq
where $\Delta \Gamma \equiv \Gamma_H - \Gamma_L$. Here $h(x)$ decreases
monotonically from 1 at $x=0$ to $1/4$ as $x \to \infty$; it is about 0.54 for
$m_t = 180$ GeV/$c^2$. In view of the sign in Eq.~(\ref{eqn:widthdiff}) and
since $\Delta m >0$ by definition, we then identify $B_s^L = B_s^{(+)}$ and
$B_s^H = B_s^{(-)}$ \cite{IsiBs}. If the mass difference $\Delta m$ turns out
to be too large to measure at present because of the rapid frequency of $B_s -
\bar B_s$ oscillations it entails, the width difference $\Delta \Gamma$ may
be large enough to detect. The possibility of a value of $\Delta \Gamma /
\Gamma$ for strange $B$ mesons large enough to measure experimentally has been
stressed previously \cite{CPeven,Blifes,largedg}.

One can measure $\bar \Gamma$ using semileptonic decays, while the decays to CP
eigenstates can be measured by studying the correlations between the
polarization states of the vector mesons in $B_s^{(\pm)} \to J/\psi \phi$. (For
similar methods applied to decays of other spinless mesons see, e.g.,
Ref.~\cite{Nelson}.)  In the present note we describe a means by which the
$J/\psi \phi$ final states of definite CP in $B_s$ decays may be separated from
one another using a simple angular distribution based on a {\it transversity}
variable \cite{PSLAC,PQUINN,IsiT}.  This transversity variable allows one to
directly separate the summed contribution of the even partial waves (S,~D) from
the odd one (P) by means of their opposite parities. The CDF Collaboration
\cite{CDF} has recently reported the first angular distribution analysis of the
decay $B_s \to J/\psi \phi$, obtaining a separation into longitudinal and
transverse helicity amplitudes without making a statement yet about the CP-even
and CP-odd contributions.

We summarize our main result.  Consider the final state $J/\psi \phi \to \ell^+
\ell^- K^+ K^-$, where $\ell = e$ or $\mu$. In the rest frame of the $J/\psi$
let the direction of the $\phi$ define the $x$ axis. Let the plane of the $K^+
K^-$ system define the $y$ axis, with $p_y(K^+) > 0$, so the normal to that
plane defines the $z$ axis.  (We assume a right-handed coordinate system.)
We define the angle $\theta$ as the angle between the
$\ell^+$ and the $z$ axis.  Then the time-dependent rate for the $J/\psi \phi$
mode is given by
$$
\frac{d^2 \Gamma}{d \cos \theta~dt} = \frac{3}{8} p(t) \left( 1 + \cos^2 \theta
\right) + \frac{3}{4} m(t) \sin^2 \theta
$$
\beq \label{eqn:dist}
= \frac{3}{8} \left[ p(t) + 2 m(t)  \right]
+ \frac{3}{8} \left[ p(t) - 2 m(t)  \right] \cos^2 \theta~~~ ,
\eeq
where
\beq
p(t) = p(0) e^{-\Gamma_L t}~~~{\rm (CP~even)}~~~,~~~
m(t) = m(0) e^{-\Gamma_H t}~~~{\rm (CP~odd)}~~~,
\eeq
so that the probability of having a CP-even [CP-odd] state at proper time $t$
is given by $p(t)/(p(t) + m(t))$ [$m(t)/(p(t) + m(t))$]. The angular
distribution is normalized in such a way that
\beq
\frac{d \Gamma}{dt} =
\int_{-1}^1 d(\cos \theta) \frac{d^2 \Gamma}{d \cos \theta~dt} = p(t)+m(t)~~~.
\eeq
As $t$ increases, one should see a growth of the $\sin^2 \theta$ component.
The angle $\theta$ is an example of a transversity variable, whose utility for
the determination of CP properties of multi-particle systems was pointed out
some time ago \cite{Bohr}.

The zero-angular-momentum states of two massive neutral vector mesons such as
$J/\psi$ and $\phi$, both with the same CP (in this case, even) consist of two
with even CP and one with odd CP.  One can form states with orbital angular
momenta $L = 0$ (CP even), $L = 1$ (CP odd), and $L = 2$ (CP even).

Alternatively, one can decompose the decay amplitude $A$ into three independent
components \cite{FM}, corresponding to linear polarization states of the vector
mesons which are either longitudinal (0), or transverse to their directions of
motion and parallel ($\parallel$) or perpendicular ($\perp$) to one another.
The states $0$ and $\parallel$ are P-even, while the state $\perp$ is P-odd.
Since $J/\psi$ and $\phi$ are both C-odd eigenstates, the properties under P
are the same as those under CP.

Consider the polarization three-vectors $\bep_{J/\psi}$ and $\bep_\phi$ in the
$J/\psi$ rest frame.  The independent decay amplitudes are the rotationally
invariant quantities linear in $\bep_{J/\psi}^*$ and $\bep_\phi^*$ and
involving
possible powers of $\bph$, a unit vector in the direction of the momentum of
$\phi$ in the $J/\psi$ rest frame.

The two CP-even decay amplitudes are the combinations $\bep^*_{J/\psi} \cdot
\bep^*_\phi$ (contributing to $A_0$ and $A_\parallel$) and $\bep^*_{J/\psi}
\cdot \bph~\bep^*_\phi \cdot \bph = \bep^{*L}_{J/\psi} \bep^{*L}_{\phi}$
(contributing only to $A_0$), where $\bep^L \equiv \bph \cdot \bep$.
Equivalently, one can subtract off the longitudinal component of the
polarization vectors to replace $\bep^*_{J/\psi} \cdot \bep^*_\phi$ by
$\bep^{*T}_{J/\psi} \cdot \bep^{*T}_\phi$, contributing only to $A_\parallel$,
where the superscripts $T$ refer to projections perpendicular to $\bph$. The
CP-odd amplitude $\bep^*_{J/\psi} \times \bep^*_\phi \cdot \bph$ contributes
only to $A_\perp$. The case of transverse ($\parallel$ or $\perp$) polarization
states is reminiscent of photon polarization correlations \cite{Yang} in
neutral pion decay. Thus we may write the decay amplitude as
\beq \label{eqn:adef}
A(B_s \to J/\psi \phi) = A_0 (m_\phi/E_\phi) \bep^{*L}_{J/\psi}
\bep^{*L}_\phi - A_\parallel \bep^{*T}_{J/\psi} \cdot \bep^{*T}_\phi/\sqrt{2}
- i A_\perp \bep^*_{J/\psi} \times \bep^*_\phi \cdot \bph/\sqrt{2}~~~,
\eeq
where $E_\phi$ is the energy of
the $\phi$ in the $J/\psi$ rest frame, and the individual amplitudes are real
in
the absence of final-state interactions.  The amplitudes for the corresponding
decays of $\bar B_s \equiv CP(B_s)$ are $\bar A_0 = A_0$, $\bar A_\parallel =
A_\parallel$, and $\bar A_\perp = - A_\perp$.  (We can see directly by counting
powers of $\bph$ that $A_0$ and $A_\parallel$ are P-even while $A_\perp$ is
P-odd.)  We have normalized the partial widths for the three independent
polarization states in such a way that
\beq
d \Gamma(B_s \to J/\psi \phi)/dt = |A_0|^2 + |A_\parallel|^2 + |A_\perp|^2~~~,
\eeq
and we may identify
\beq
p(t) = |A_0|^2 + |A_\parallel|^2~~~,~~~m(t) = |A_\perp|^2~~~.
\eeq
For purposes of comparing with other notations, we can express the helicity
amplitudes $A_\lambda$ (where $\lambda = 1,~0,~-1$ is the projection of the
$\phi$ angular momentum on the $x$ axis) in terms of the linear polarization
basis by $A_{\pm 1} = (A_\parallel \pm A_\perp)/ \sqrt{2}$, with $A_0$ the same
in either basis, and in terms of S-, P-, and D-wave amplitudes by
\beq
A_{\pm 1} = \sqrt{\frac{1}{3}} S \pm \sqrt{\frac{1}{2}} P
+ \sqrt{\frac{1}{6}} D~~~,~~~
A_0 = - \sqrt{\frac{1}{3}} S + \sqrt{\frac{2}{3}} D~~~.
\eeq
With these normalizations,
\beq
d \Gamma(B_s \to J/\psi \phi)/dt = |A_0|^2 + |A_1|^2 + |A_{-1}|^2
= |S|^2 + |P|^2 + |D|^2~~~,
\eeq
and
\beq
A_\parallel = \sqrt{\frac{2}{3}} S + \sqrt{\frac{1}{3}} D~~~,~~~
A_\perp = P~~~.
\eeq
The longitudinal and transverse partial widths are given, respectively, by
\beq
d \Gamma_0/dt = |A_0|^2~~~,~~~d \Gamma_T/dt = |A_1|^2 + |A_{-1}|^2~~~.
\eeq
In terms of partial-wave amplitudes, one has
$$
\frac{d \Gamma_0}{dt} = |- \sqrt{1/3} S + \sqrt{2/3} D|^2 ~~~,~~~
\frac{d \Gamma_T}{dt} = |\sqrt{2/3} S +\sqrt{1/3}D|^2 +|P|^2~~~,
$$
\beq
\frac{d \Gamma_\parallel}{dt} = |\sqrt{2/3} S +\sqrt{1/3}D|^2~~~,
\frac{d \Gamma_\perp}{dt} = |P|^2~~~,
\eeq
while
\beq
\frac{p(t)}{p(t) + m(t)} = \frac{|S|^2 + |D|^2}{|S|^2 + |P|^2 + |D|^2}~~~;~~~
\frac{m(t)}{p(t) + m(t)} = \frac{|P|^2}{|S|^2 + |P|^2 + |D|^2}~~~.
\eeq
Finally, we note that in the covariant expression ~\cite{KP}
\beq
A_\lambda = \epsilon_{1 \mu}^* \epsilon_{2 \nu}^* \left[ a g^{\mu
\nu} + \frac{b}{m_1 m_2} p_2^\mu p_1^\nu + \frac{i c}{m_1 m_2} \epsilon^{\mu
\nu \alpha \beta} p_{1 \alpha} p_{2 \beta} \right]~~~
\eeq
for the decay $B \to V_1 V_2$, where $\epsilon^{0123} \equiv +1$ and
$V_1$ and $V_2$ are vector mesons with
masses $m_1$ and $m_2$ and four-momenta $p_1$ and $p_2$, the helicity
amplitudes are
\beq
A_{\pm 1} = a \pm c \sqrt{x^2-1}~~~,~~~
A_0 = - ax - b(x^2-1)~~~,
\eeq
where $x \equiv p_1 \cdot p_2/(m_1 m_2)$.  We thus identify
\beq
S = \frac{1}{\sqrt{3}} \left[ a(2+x) + b(x^2 -1) \right]~~,~~
P = c \sqrt{2(x^2-1)}~~,~~
D = \sqrt{\frac{2}{3}} \left[ a(1-x) - b(x^2-1) \right]~~.
\eeq
Note that $S$ and $D$ both involve $a$ and $b$.

The derivation of Eq.~(\ref{eqn:dist}) is elementary.  The $\phi$ is coupled to
$K^+ K^-$ through an amplitude $\epsilon_\phi \cdot (p_{K^+} - p_{K^-})$, where
the quantities denote 4-vectors.  Thus the plane of (linear) $\phi$
polarization is related to that of the $K^+ K^-$ system in the $J/\psi$ rest
frame.  By definition, we have taken the $\phi$ linear polarization vector to
lie in the $x-y$ plane.  We may define an angle $\psi$ as that of the $K^+$ in
the $\phi$ rest frame relative to the helicity axis (the negative of the
direction of the $J/\psi$ in that frame).
The spatial components of the $\phi$ and $J/\psi$ polarizations must be
correlated since the decaying strange $B$ is spinless. The $J/\psi$ then has a
single linear polarization state $\bep$ for each amplitude: In the $J/\psi$
rest frame,
\beq
A_\parallel:~~~\bep = \hat y~~~;~~~
A_0        :~~~\bep = \hat x~~~;~~~
A_\perp    :~~~\bep = \hat z~~~.
\eeq
A unit vector $n$ in the direction of the $\ell^+$
in $J/\psi$ decay may be defined to have components
\beq \label{eqn:ndef}
(n_x, n_y, n_z) = (\sin \theta \cos \varphi,
\sin \theta \sin \varphi, \cos \theta)~~~
\eeq
where $\varphi$ is the angle between the projection of the $\ell^+$ on the $K^+
K^-$ plane in the $J/\psi$ rest frame and the $x$ axis.  The sum over lepton
polarizations then leads to a tensor in the $J/\psi$ rest frame with spatial
components (in the limit of zero lepton mass, assumed here)
\beq \label{eqn:lijdef}
\sum_{\ell^\pm {\rm pol}} [\bar u \gamma_i v]^* [\bar u\gamma_j v]
\sim L_{ij} \equiv \delta_{ij} - n_i n_j~~~.
\eeq
Physically this tensor simply expresses the fact that massless lepton pairs
couple only to transverse polarization states of the $J/\psi$, as expected from
the structure of the electromagnetic interactions.

Taking account of the definition (\ref{eqn:adef}), we then find
that the probability for the decay $B_s \to (\ell^+\ell^-)_{J/\psi}
(K^+K^-)_{\phi}$ is proportional to
\beq
\sum_{\ell^\pm {\rm pol}} |A|^2 = A_i A^*_j L_{ij}~~~,
\eeq
where
\beq
A_i = A_0 \delta_{ix} \cos \psi - A_\parallel \delta_{iy} \sin \psi/\sqrt{2}
+ i A_\perp \delta_{iz} \sin \psi/\sqrt{2}~~~.
\eeq
Consequently, when we use the definitions (\ref{eqn:ndef}), we find
$$
\frac{d^4 \Gamma [B_s \to (\ell^+\ell^-)_{J/\psi} (K^+K^-)_{\phi}]}
{d \cos \theta~d \varphi~d \cos \psi~dt}
= \frac{9}{32 \pi} [2 |A_0|^2 \cos^2 \psi (1 - \sin^2 \theta \cos^2 \varphi)
$$
$$
+ \sin^2 \psi \{ |A_\parallel|^2  (1 - \sin^2 \theta \sin^2 \varphi)
+ |A_\perp|^2 \sin^2 \theta - {\rm Im}(A_\parallel^* A_\perp)
\sin 2 \theta \sin \varphi \}~~~
$$
\beq \label{eqn:threeangle}
+\frac{1}{\sqrt{2}}\sin 2 \psi \{{\rm Re}(A_0^* A_\parallel) \sin^2 \theta
\sin 2 \varphi + {\rm Im}(A_0^* A_\perp) \sin 2 \theta \cos \varphi \} ]~~~.
\eeq
The overall normalization has been chosen to agree with our previous
conventions when one integrates over angles.  For $\bar B_s$ decays
the interference terms involving $A_\perp$ amplitudes are of
opposite sign and all other terms are unchanged.

Integration over $\cos \psi$ leads to the distribution
$$
\frac{d^3 \Gamma [B_s \to (\ell^+\ell^-)_{J/\psi} (K^+K^-)_{\phi}]}
{d \cos \theta~d \varphi~dt}
= \frac{3}{8 \pi} [|A_0|^2 (1 - \sin^2 \theta \cos^2 \varphi)
+ |A_\parallel|^2 (1 - \sin^2 \theta \sin^2 \varphi)
$$
\beq \label{eqn:twoangle}
+ |A_\perp|^2 \sin^2 \theta - {\rm Im}~(A_\parallel^* A_\perp)
\sin 2 \theta \sin \varphi ]~~~.
\eeq
Performing the integrals over $\varphi$ and taking account of the differing
time-dependences of the decays of $B_s^{(\pm)}$, we obtain the result
(\ref{eqn:dist}).  This is a suitable single-angle distribution to employ if
one wishes to disentangle the CP-even and CP-odd components of the $B_s$. The
two-angle distribution (\ref{eqn:twoangle}) allows one to separate out the
individual quantities $|A_0|^2$, $|A_\parallel|^2$, and $|A_\perp|^2$.

An interesting oscillation appears in the interference terms between CP-even
and CP-odd decays.  For example, in the two-angle distribution
(\ref{eqn:twoangle}), since the respective time-dependences of the $B_s^{(+)}$
and $B_s^{(-)}$ decay amplitudes are $e^{-i m_L t - \Gamma_L t/2}$ and $e^{-i
m_H t - \Gamma_H t/2}$, the term $ - {\rm Im}~(A_\parallel^* A_\perp)$ behaves
as $ |A_\parallel(0) A_\perp(0)| \sin(\Delta m t - \delta) e^{-\overline \Gamma
t}$, where $\delta$ is a strong final-state phase shift difference:
$A_\parallel(0)^* A_\perp(0) = |A_\parallel(0) A_\perp(0)| e^{i \delta}$. Thus,
if one tags the flavor of the decaying $B_s$, one can observe the effects of
$\Delta m$ in the $J/\psi \phi$ final state through the interference of final
states of opposite CP (if both are present). The oscillation term averages out
to zero if the initial numbers of $B_s$ and $\bar B_s$ are equal.

The distributions (\ref{eqn:dist}), (\ref{eqn:threeangle}), and
(\ref{eqn:twoangle}) also permit one to separate out the components
$|A_0|^2$, $|A_\parallel|^2$, $|A_\perp|^2$, and the interference terms ${\rm
Im}(A_\parallel^* A_\perp)$, ${\rm Re}(A_0^* A_\parallel) $, and ${\rm
Im}(A_0^* A_\perp)$ for the decays $B^0 \to J/\psi K^{*0}$.  (Here and
subsequently we imply the sum over a process and its charge-conjugate.)
Moreover, in the
limit of flavor SU(3) symmetry, one expects the ratios of the relative
components in $B^0 \to J/\psi K^{*0}$ to be the same as those {\it at
proper time $t=0$} in the decays $B_s \to J/\psi \phi$ \cite{Isi93}.
Thus, an analysis of
$B^0 \to J/\psi K^{*0}$ can provide an independent estimate of the relative
contributions of CP-even and CP-odd final states at $t=0$ to the decays $B_s
\to J/\psi \phi$, enhancing the ability to determine $\Gamma_H$ and $\Gamma_L$.

If the $K^*$ is observed to decay to the CP eigenstate $K_S \pi^0$, the
amplitudes $A_0$ and $A_\parallel$ refer (as in the case of $J/\psi \phi$) to
the CP-even eigenstate, while $A_\perp$ refers to the CP-odd eigenstate. The
expected dominance of the CP-even eigenstate (see below) means that in $B^0 \to
J/\psi K_S \pi^0$ events the CP asymmetry will tend to be opposite to that in
$B^0 \to J/\psi K_S$ \cite{PQUINN,KKPS}.
Since the rates for observing the processes
$B^0 \to J/\psi K_S$ and $B^0 \to J/\psi K^{*0} \to J/\psi K_S \pi^0$
are comparable (taking account of branching ratios
and typical detection efficiencies), the incorporation of
$J/\psi K_S \pi^0$ data may add statistical power to any experiment studying
the $J/\psi K_S$ final state, even when the $\pi^0$ is not observed directly
but its existence inferred.

The distribution (\ref{eqn:dist}) permits one to separate amplitudes of
opposite parity from one another even if the $K^+K^-$ system in $B_s \to J/\psi
 K^+ K^-$ or the $K \pi$ system in $B \to J/\psi K \pi$ is not a vector meson
\cite{PQUINN}.  This is easily seen by considering the density matrix
$\rho_{ij}$ of the $J/\psi$, expressed in terms of linear polarization states,
so that the decay rate is proportional to $\rho_{ij} L_{ij}$, with $L_{ij}$
defined in (\ref{eqn:lijdef}).  If we integrate over $\varphi$, we find
\beq
\frac{d^2\Gamma}{d \cos \theta~dt} \sim (\rho_{xx} + \rho_{yy}) \left( \frac{1
+ \cos^2 \theta}{2} \right) + \rho_{zz} \sin^2 \theta~~~,
\eeq
where $\rho_{xx}$ and $\rho_{yy}$ correspond to linear $J/\psi$ polarization
states in the plane of the two pseudoscalar mesons, while $\rho_{zz}$
corresponds to $J/\psi$ polarization perpendicular to this plane, and thus
represents an amplitude with parity opposite to those contributing to
$\rho_{xx}$ and $\rho_{yy}$. For cases where each particle in
the final state is a C eigenstate
as in $B^0 \to J/\psi K^{*0} \to J/\psi K_S \pi^0$  and $B_s \to J/\psi \phi
\to J/\psi K_S K_L$ the parity separation is also a CP separation and the
transversity analysis can be used without the need to extract the vector
resonance from nonresonant or other background.

The dominance of the $|A_0|^2$ contribution in $B^0 \to J/\psi K^{*0}$ decays
\cite{CDF,ARGUS,CLEO} implies via flavor SU(3) that the $|A_0|^2$ contribution
should also dominate $B_s \to J/\psi \phi$, and hence that $B_s^{(-)} \to
J/\psi
\phi$ is likely to be suppressed in comparison with $B_s^{(+)} \to J/\psi
\phi$.  Thus the initial angular distribution is very likely to be dominated by
the $1 + \cos^2 \theta$ component. As time increases, the fraction of the
angular distribution proportional to this component will decrease while that
proportional to $\sin^2 \theta$ will increase.  It should be possible to
separate out the two components by a combined analysis in $\theta$ and proper
decay time.  If the $\sin^2 \theta$ component does not show up even at large
times, a single-exponential fit to the decay should provide a good estimate of
the lifetime of the CP-even eigenstate.

The angular distribution in (\ref{eqn:dist}) has the form of an
ellipsoid which is prolate if $p(t) > 2 m(t)$ and oblate if $p(t) < 2 m(t)$.
(Here we imagine an average over $\varphi$ to have been performed.) If
all three partial waves are equally populated $p(t) = 2 m(t)$ since there are
two partial waves with even CP and only one with odd CP. For this case the
angular distribution is isotropic in $\cos \theta$ as expected.

If $p(0) > 2 m(0)$ (i.e., if the CP-even decay is initially more than 2/3
dominant), if the CP-even eigenstate $B_s^{(+)}$ has the greater decay rate
as expected, and if there is a non-zero odd-CP component $m(0) \ne 0$, then
the angular distribution in transversity will be initially prolate
but will eventually become oblate as the quantity $p(t) - 2 m(t)$ changes sign.
This effect can be noted by dividing the events into two bins with $|\cos
\theta| < 1/2$ and $|\cos \theta| > 1/2$, denoted by $E$ (equatorial) and $P$
(polar) respectively:

\beq
E \equiv \int_{-1/2}^{1/2} d(\cos \theta) \frac{d^2 \Gamma}{d \cos \theta~dt}
= \frac{13}{32} p(t) +  \frac{11}{16} m(t)~~~,
\eeq

\beq
P \equiv \left[ \int_{-1}^{-1/2} + \int_{1/2}^1 \right]
d(\cos \theta) \frac{d^2 \Gamma}{d \cos \theta~dt}
= \frac{19}{32} p(t) + \frac{5}{16} m(t)~~~,
\eeq

\beq
P - E = \frac{3}{16} \left[ p(t) - 2 m(t)  \right]
\eeq

The difference between the numbers in the two bins provides an experimental
number whose sign will change in time under the assumptions noted above.
This two-bin analysis of data can be adjusted for optimum statistics by
changing the sizes of the bins to correspond to the initial values of $p(0)$
and $m(0)$.

The analysis performed by CDF \cite{CDF} for $B_s \to J/\psi \phi$ was also
based on a single-angle distribution, but it separated out the transverse
component from the longitudinal component.  In the absence of vertex cuts these
would be, respectively,
$\Gamma_T \equiv \int_0^\infty dt (|A_\parallel|^2 + |A_\perp|^2)$
and $\Gamma_0 \equiv \int_0^\infty dt (|A_0|^2)$.  With a minimum vertex cut
of 50 $\mu$m, the result obtained was
$\Gamma_0/(\Gamma_0 + \Gamma_T) = 0.56 \pm 0.21
{\rm~(stat)}_{~-0.04}^{~+0.02}{\rm~(sys)}$. The transverse component contains
both CP-even and CP-odd contributions, while the longitudinal component is
CP-even.  The separation of transverse and longitudinal components makes sense
only if the $B_s^{(-)} \to J/\psi \phi$ decay is negligible, or if the lifetime
difference between $B_s^{(+)}$ and $B_s^{(-)}$ can be ignored.

Corresponding determinations of $\Gamma_0/(\Gamma_0 + \Gamma_T)$ for the
decay $B^0 \to J/\psi K^{*0}$ are
$0.65 \pm 0.10 \pm 0.04$ (CDF) \cite{CDF},
$0.97 \pm 0.16 \pm 0.15$ (ARGUS) \cite{ARGUS},
$0.80 \pm 0.08 \pm 0.05$ (CLEO) \cite{CLEO}, and
$0.74 \pm 0.07$ (world average) \cite{CDF}.  This last value is compatible with
the corresponding one for $B_s \to J/\psi \phi$. A discrepancy would
have indicated either a violation of SU(3) or the lifetime effect mentioned
above.

The presence of two eigenstates with possibly differing lifetimes can affect
any determination of $\tau(B_s)$.  When observing the $B_s$ in a final state of
definite flavor, such as $D_s \pi$ or $D_s \ell \nu_\ell$, one will be
effectively measuring the average lifetime $\bar \Gamma$ of the CP-even and
CP-odd states. Most measurements reported up to now, including a recent CDF
determination \cite{CDFBs} leading to a world average \cite{Kroll} of
$\tau(B_s) = 1.58 \pm 0.10$ ps, are of this type.  This quantity is expected
\cite{Blifes} to be very close to the $B^0$ lifetime, for which the world
average \cite{Kroll} is $\tau(B^0) = 1.57 \pm 0.05$ ps. However,
minimum-lifetime cuts can bias the sample against the CP-even (shorter-lived)
component, leading to results which depend on the cut if a single-exponential
fit is adopted.

To summarize, we have found that a combined analysis with respect to proper
decay time and a single transversity angle in the decay $B_s \to J/\psi
\phi$ can determine the lifetime of at least the CP-even and possibly the
CP-odd mass eigenstates of the $B_s - \bar B_s$ system.  Additional
information about the properties of the $J/\psi \phi$ mode at proper time $t =
0$ can be obtained by a similar analysis of the decays $B^0 \to J/\psi K^{*0}$.
These analyses can already be attempted with the data sample \cite{CDF} now in
hand.

We thank B. Kayser, R. Kutschke,
K. Ohl, M. Paulini, E. A. Paschos, M. P. Schmidt,
and W. Wester for useful discussions. This work was supported in part by the
United States Department of Energy under Contracts No. DE AC02 76CH03000,
DE FG02 90ER40560 and W-31-109-ENG-38.
\newpage

\def \ajp#1#2#3{Am.~J.~Phys.~{\bf#1} (#3) #2}
\def \apny#1#2#3{Ann.~Phys.~(N.Y.) {\bf#1} (#3) #2}
\def \app#1#2#3{Acta Phys.~Polonica {\bf#1} (#3) #2 }
\def \arnps#1#2#3{Ann.~Rev.~Nucl.~Part.~Sci.~{\bf#1} (#3) #2}
\def \cmp#1#2#3{Commun.~Math.~Phys.~{\bf#1} (#3) #2}
\def \cmts#1#2#3{Comments on Nucl.~Part.~Phys.~{\bf#1} (#3) #2}
\def \cn{Collaboration}
\def \corn93{{\it Lepton and Photon Interactions:  XVI International Symposium,
Ithaca, NY August 1993}, AIP Conference Proceedings No.~302, ed.~by P. Drell
and D. Rubin (AIP, New York, 1994)}
\def \cp89{{\it CP Violation,} edited by C. Jarlskog (World Scientific,
Singapore, 1989)}
\def \dpff{{\it The Fermilab Meeting -- DPF 92} (7th Meeting of the American
Physical Society Division of Particles and Fields), 10--14 November 1992,
ed. by C. H. Albright \ite~(World Scientific, Singapore, 1993)}
\def \dpf94{DPF 94 Meeting, Albuquerque, NM, Aug.~2--6, 1994}
\def \efi{Enrico Fermi Institute Report No. EFI}
\def \el#1#2#3{Europhys.~Lett.~{\bf#1} (#3) #2}
\def \f79{{\it Proceedings of the 1979 International Symposium on Lepton and
Photon Interactions at High Energies,} Fermilab, August 23-29, 1979, ed.~by
T. B. W. Kirk and H. D. I. Abarbanel (Fermi National Accelerator Laboratory,
Batavia, IL, 1979}
\def \hb87{{\it Proceeding of the 1987 International Symposium on Lepton and
Photon Interactions at High Energies,} Hamburg, 1987, ed.~by W. Bartel
and R. R\"uckl (Nucl. Phys. B, Proc. Suppl., vol. 3) (North-Holland,
Amsterdam, 1988)}
\def \ib{{\it ibid.}~}
\def \ibj#1#2#3{~{\bf#1} (#3) #2}
\def \ichep72{{\it Proceedings of the XVI International Conference on High
Energy Physics}, Chicago and Batavia, Illinois, Sept. 6--13, 1972,
edited by J. D. Jackson, A. Roberts, and R. Donaldson (Fermilab, Batavia,
IL, 1972)}
\def \ijmpa#1#2#3{Int.~J.~Mod.~Phys.~A {\bf#1} (#3) #2}
\def \ite{{\it et al.}}
\def \jmp#1#2#3{J.~Math.~Phys.~{\bf#1} (#3) #2}
\def \jpg#1#2#3{J.~Phys.~G {\bf#1} (#3) #2}
\def \lkl87{{\it Selected Topics in Electroweak Interactions} (Proceedings of
the Second Lake Louise Institute on New Frontiers in Particle Physics, 15--21
February, 1987), edited by J. M. Cameron \ite~(World Scientific, Singapore,
1987)}
\def \ky85{{\it Proceedings of the International Symposium on Lepton and
Photon Interactions at High Energy,} Kyoto, Aug.~19-24, 1985, edited by M.
Konuma and K. Takahashi (Kyoto Univ., Kyoto, 1985)}
\def \mpla#1#2#3{Mod.~Phys.~Lett.~A {\bf#1} (#3) #2}
\def \nc#1#2#3{Nuovo Cim.~{\bf#1} (#3) #2}
\def \np#1#2#3{Nucl.~Phys.~{\bf#1} (#3) #2}
\def \pisma#1#2#3#4{Pis'ma Zh.~Eksp.~Teor.~Fiz.~{\bf#1} (#3) #2[JETP Lett.
{\bf#1} (#3) #4]}
\def \pl#1#2#3{Phys.~Lett.~{\bf#1} (#3) #2}
\def \plb#1#2#3{Phys.~Lett.~B {\bf#1} (#3) #2}
\def \pr#1#2#3{Phys.~Rev.~{\bf#1} (#3) #2}
\def \pra#1#2#3{Phys.~Rev.~A {\bf#1} (#3) #2}
\def \prd#1#2#3{Phys.~Rev.~D {\bf#1} (#3) #2}
\def \prl#1#2#3{Phys.~Rev.~Lett.~{\bf#1} (#3) #2}
\def \prp#1#2#3{Phys.~Rep.~{\bf#1} (#3) #2}
\def \ptp#1#2#3{Prog.~Theor.~Phys.~{\bf#1} (#3) #2}
\def \rmp#1#2#3{Rev.~Mod.~Phys.~{\bf#1} (#3) #2}
\def \rp#1{~~~~~\ldots\ldots{\rm rp~}{#1}~~~~~}
\def \si90{25th International Conference on High Energy Physics, Singapore,
Aug. 2-8, 1990}
\def \slc87{{\it Proceedings of the Salt Lake City Meeting} (Division of
Particles and Fields, American Physical Society, Salt Lake City, Utah, 1987),
ed.~by C. DeTar and J. S. Ball (World Scientific, Singapore, 1987)}
\def \slac89{{\it Proceedings of the XIVth International Symposium on
Lepton and Photon Interactions,} Stanford, California, 1989, edited by M.
Riordan (World Scientific, Singapore, 1990)}
\def \smass82{{\it Proceedings of the 1982 DPF Summer Study on Elementary
Particle Physics and Future Facilities}, Snowmass, Colorado, edited by R.
Donaldson, R. Gustafson, and F. Paige (World Scientific, Singapore, 1982)}
\def \smass90{{\it Research Directions for the Decade} (Proceedings of the
1990 Summer Study on High Energy Physics, June 25 -- July 13, Snowmass,
Colorado), edited by E. L. Berger (World Scientific, Singapore, 1992)}
\def \smassb{{\it Proceedings of the Workshop on $B$ Physics at Hadron
Accelerators}, Snowmass, Colorado, June 21 -- July 2, 1993, edited by
P. McBride and C. S. Mishra (Fermilab report Fermilab-CONF-93/267, 1993)}
\def \stone{{\it B Decays}, edited by S. Stone (World Scientific, Singapore,
1994)}
\def \tasi90{{\it Testing the Standard Model} (Proceedings of the 1990
Theoretical Advanced Study Institute in Elementary Particle Physics, Boulder,
Colorado, 3--27 June, 1990), edited by M. Cveti\v{c} and P. Langacker
(World Scientific, Singapore, 1991)}
\def \yaf#1#2#3#4{Yad.~Fiz.~{\bf#1} (#3) #2 [Sov.~J.~Nucl.~Phys.~{\bf #1} (#3)
#4]}
\def \zhetf#1#2#3#4#5#6{Zh.~Eksp.~Teor.~Fiz.~{\bf #1} (#3) #2 [Sov.~Phys. -
JETP {\bf #4} (#6) #5]}
\def \zpc#1#2#3{Zeit.~Phys.~C {\bf#1} (#3) #2}


\begin{thebibliography}{99}

\bibitem{CKM} N. Cabibbo, \prl{10}{531}{1963}; M. Kobayashi and T. Maskawa,
\ptp{49}{652}{1973}.

\bibitem{JRCP} See, e.g., J. L. Rosner, \efi~95-36, hep-ph/9506364, lectures
presented at the VIII J. A. Swieca Summer School, Rio de Janeiro, Feb.~7--11,
1995, proceedings to be published by World Scientific.

\bibitem{CPeven} R. Aleksan, A. Le Yaouanc, L. Oliver, O. P\`ene, and J. C.
Raynal, \plb{316}{567}{1993}.

\bibitem{Blifes} M. B. Voloshin, N. G. Uraltsev, V. A. Khoze, and M. A.
Shifman, \yaf{46}{181}{1987}{112}; I. I. Bigi \ite, in \stone, p.~132;
I. I. Bigi, Univ. of Notre Dame Report No.~UND-HEP-95-BIG02, June, 1995,
to be published in Physics Reports.

\bibitem{Soni} A. Soni, Brookhaven National Laboratory report BNL-62284,
October, 1995, to be published in Proceedings of Lattice 95, Melbourne,
Australia, July 10-13, 1995.

\bibitem{IsiBs} I. Dunietz, \prd{52}{3048}{1995}.

\bibitem{BDY} G. Buchalla, I. Dunietz, and H. Yamamoto, FERMILAB-PUB-95/167-T,
July, 1995, to be published in Phys.~Lett.~B.

\bibitem{BP} T. E. Browder and S. Pakvasa, \prd{52}{3123}{1995}.

\bibitem{largedg} L. L. Chau, W.-Y. Keung, and M. D. Tran,
\prd{27}{2145}{1983}; L. L. Chau, \prp{95}{1}{1983};
A. J. Buras, W. Slominsky, and H. Steger, \np{B245}{369}
{1984}; A. Datta, E. A. Paschos, and U. T\"urke, \plb{196}{382}
{1987}; A. Datta, E. A. Paschos, and Y. L. Wu, \np{B311}{35}{1988}; I. Dunietz,
\apny{184}{350}{1988}.

\bibitem{Nelson} N.-P. Chang and C. A. Nelson, \prl{40}{1617}{1978};
\prd{20}{2923}{1978}; C. A. Nelson, \prd{30}{107, 1937}{1984};
\ibj{32}{1848(E)}{1985}.

\bibitem{PSLAC} H. J. Lipkin, in {\it Proceedings of the SLAC Workshop on
Physics and Detector Issues for a High-Luminosity Asymmetric B Factory}, edited
by David Hitlin, published as SLAC, LBL and Caltech reports SLAC-373, LBL-30097
and CALT-68-1697, 1990, p. 49.

\bibitem{PQUINN} I. Dunietz, H. R. Quinn, A. Snyder, W. Toki and H. J. Lipkin,
\prd{43}{2193}{1991}.

\bibitem{IsiT} I. Dunietz, Appendix in \stone, p.~550.

\bibitem{CDF} CDF \cn, F. Abe \ite, \prl{75}{3068}{1995}.

\bibitem{Bohr} A. Bohr, \np{10}{486}{1959}.

\bibitem{FM} J. L. Rosner, \prd{42}{3732}{1990}.

\bibitem{Yang}  L. D. Landau, Dokl.~Akad.~Nauk SSSR {\bf 60} (1948) 207;
C. N. Yang, \pr{77}{242, 722}{1950}.

\bibitem{KP} G. Kramer and W. F. Palmer, \prd{45}{193}{1992}; \ibj{46}
{2969, 3197}{1992}; \zpc{55}{497}{1992}.

\bibitem{Isi93} I. Dunietz, in \smassb, p.~83.

\bibitem{KKPS} B. Kayser, M. Kuroda, R. D. Peccei, and A. I. Sanda,
\plb{237}{508}{1990}.

\bibitem{ARGUS} ARGUS \cn, H. Albrecht \ite, \plb{340}{217}{1994}.

\bibitem{CLEO} CLEO \cn, M. S. Alam \ite, \prd{50}{43}{1994}.

\bibitem{CDFBs} CDF \cn, F. Abe \ite, \prl{74}{4988}{1995}.

\bibitem{Kroll} I. J. Kroll, presented at 17th International Symposium on
Lepton and Photon Interactions, Beijing, 11 August 1995.

\end{thebibliography}
\end{document}